# High contrast few-cycle frontend with hybrid amplification for petawatt-class lasers

Roland S. Nagymihály,[1,*] Mikhail Kalashnikov,[1] Levente Lehotai,[1] Viktor Pajer,[1] János Bohus,[1] Nóra Csernus-Lukács,[1] János Csontos,[1] Szabolcs Tóth,[1] Balázs Tari,[1] Ignas Balciunas,[2] Ernestas Kucinskas,[2] Tomas Stanislauskas,[2] and Ádám Börzsönyi[1]

[1]ELI ALPS, ELI-HU Non-Profit Ltd., Wolfgang Sandner u. 3, 6728-Szeged, Hungary
[2]Light Conversion Ltd., Keramiku str. 2b, LT-10233 Vilnius, Lithuania
roland.nagymihaly@eli-alps.hu

**Abstract:** We present a 10 mJ-class laser system for seeding petawatt-class Ti:Sa lasers based on a hybrid, femtosecond optical parametric and negatively chirped Ti:Sa-based amplification architecture. Output pulses with 13 mJ energy, 14.4 fs transform limited duration, and high spatio-spectral quality are reached at 100 Hz repetition rate centered at 788 nm wavelength. By compressing 1 mJ energy with bulk glass and positively chirped mirrors, a pulse duration of 15.5 fs is measured. The intensity contrast is higher than $10^{12}$ already 15 ps before the main pulse. The presented architecture can be a robust and reliable solution for seeding sub-17 fs high intensity lasers with ultrahigh contrast at up to 100 Hz repetition rate.

## 1. Introduction

Recent advances in petawatt class laser systems have shown significant focus on the development of next generation frontend arrangements [1-5]. Here, we consider the components of a double chirped pulse amplification (DCPA) system [6] located before the last stretcher as a frontend. In multi cascade laser systems the first cascades determine limits of major pulse parameters as spectral bandwidth and temporal contrast that can be reached at the end of the chain. Since in power amplifiers the latter can be conserved, or degraded only, large scale laser facilities require reliable frontends, which can provide appropriate spatio-spectral properties, ultrahigh temporal contrast, high energy and ultimate long-term stability. These properties are, naturally, very challenging to reach simultaneously. Most present-day petawatt class laser systems utilize frontends with the following architecture. The scheme starts with a Ti:Sa femtosecond oscillator, amplified in a Ti:Sa-based CPA system, after which spatio-temporal cleaning is performed by a cross-polarized wave (XPW) generation stage. The output of such a frontend typically lies in the range of 50-100 μJ, while the repetition rate is 10 Hz to 1 kHz, and the spectral bandwidth covers the 740–860 nm window. The temporal intensity contrast is usually limited to $10^{10}$. Such a system is used, for example, in the BELLA [3], VEGA 3 [7], or in the HAPLS laser [8].

In the last decade, several different approaches surfaced with the utilization of optical parametric chirped pulse amplification (OPCPA). One solution comprises of a Ti:Sa-based CPA stage seeded by a Ti:Sa oscillator, after which an XPW stage improves the contrast. Then the pulse is further amplified in picosecond OPCPA stages with moderate stretching, to keep a sufficiently large spectral bandwidth by avoiding spectral narrowing associated with high gain and saturated Ti:Sa amplification. Such a frontend operates in APOLLON [1,9] and in the 10 PW HPLS laser [10]. These frontend types provide pulses with energies up to 10 mJ at 10–100 Hz, and with a spectral bandwidth covering the range of 690–910 nm in the broadest case, while the temporal contrast reaches $10^{11}$–$10^{13}$ on the pre-pulse side [1,10]. Another approach uses a supercontinuum seeded OPCPA system, which obtains the seed pulses through a difference frequency generation (DFG) and utilizes second harmonic generation (SHG) twice to reach the pulse spectrum corresponding to the Ti:Sa gain band [2]. This kind of frontend can emit pulses with an energy of 10 µJ, at a repetition rate of multi-kHz, a spectrum covering the range of 740–880 nm, and a temporal contrast of >$10^{13}$. Such a frontend is implemented in the HFPW laser of ELI ALPS [11]. A third approach has been shown to be also effective in providing a temporal contrast of >$10^{11}$ by using a combination of XPW and cascaded femtosecond OPA stages driven by a multi-mJ 40 fs Ti:Sa CPA system [5]. A novel, fully OPCPA-based frontend design was presented just recently [12], where the supercontinuum and DFG-based seed generation is separated from the power amplification section. After reaching the few-µJ level, the booster amplification is foreseen in either Ti:Sa amplifiers or OPCPA stages.

Ultrabroadband amplification in Ti:Sa was investigated to provide 0.24 TW pulses with <13 fs duration and 3.2 mJ energy in [13]. Furthermore, 0.9 TW pulses and 18.7 fs duration, but with 16 mJ energy [14], based on multipass and a combination of regenerative and multipass amplifiers. In [13], spectral management was performed by using spectral filtering mirrors in the multipass amplifier and by utilizing the acousto-optic programmable dispersive filter (AOPDF) in double-pass configuration. In case of the 0.9 TW amplifier system, spectral shaping was obtained by using an acousto-optic programmable gain filter (Mazzler, Fastlite) [14]. In both cases, the seed source was a Ti:Sa oscillator, and temporal contrast at the output of the frontend was not reported.

Here we present performances of our frontend architecture after the first development phase. Together with recent achievements with OPCPA systems, where a very short duration of the stretched and seed pulses could be realized (supporting a short pre-pedestal), amplification in Ti:Sa is much more robust and has lower requirements for the pump. Because of that, in our system we exploit advantages of both amplification schemes. This frontend can be an ideal seed source for 100 Hz multi-J, or 10s Hz multi-PW sub-17 fs amplifier systems. We would like to highlight the fact, that we did not utilize any spatio-temporal cleaning to obtain the results discussed later.

## 2. Architecture

### 2.1 Overall design

In our scheme (Fig. 1), we address the following requirements on an ideal frontend for a broadband Ti:Sa-based petawatt-class laser system: spectral bandwidth supporting sub-15 fs pulses, output energy in the >10 mJ range, ultrahigh temporal contrast, and increased robustness by utilizing Ti:Sa amplifiers with diode-pumped solid-state lasers (DPSSL) as pump sources. As a seeder, we use an industrial Yb:KGW laser (Pharos 2 mJ, 190 fs, Light Conversion) pumped femtosecond OPCPA system (customized ORPHEUS-OPCPA-800, Light Conversion). The uncompressed pulses of the OPCPA are sent through a GRISM stretcher arrangement (Light Conversion), providing negative stretching. There are two reasons for choosing negative stretching. First, since our laser is designed to be used as a seeder for high peak power Ti:Sa systems, a combination of negative stretching in the first stages and positive stretching in power amplifiers supports a broader overall bandwidth [15]. Second, if using as a standalone system, compression with bulk allows substantially lower energy and bandwidth losses than diffraction gratings. The repetition rate of the seed pulses is decreased from 1 kHz to 100 Hz by a Pockels cell, which also serves as back-scattering protection towards the OPCPA seeder. The negatively stretched pulses are shaped in spectral phase and amplitude by an AOPDF (Dazzler, Fastlite), and a polarization-encoded filter (PEF) [17]. The spectrally shaped pulses are amplified in two multipass Ti:Sa stages (Fig. 1, MPA1 and MPA2). Since diagnostics used for the pulse characterization did not require full energy, for simplicity of testing the compression, a portion of the amplified pulse energy is sampled by a reflection on a glass wedge, yielding about 1 mJ in the reflected beam.

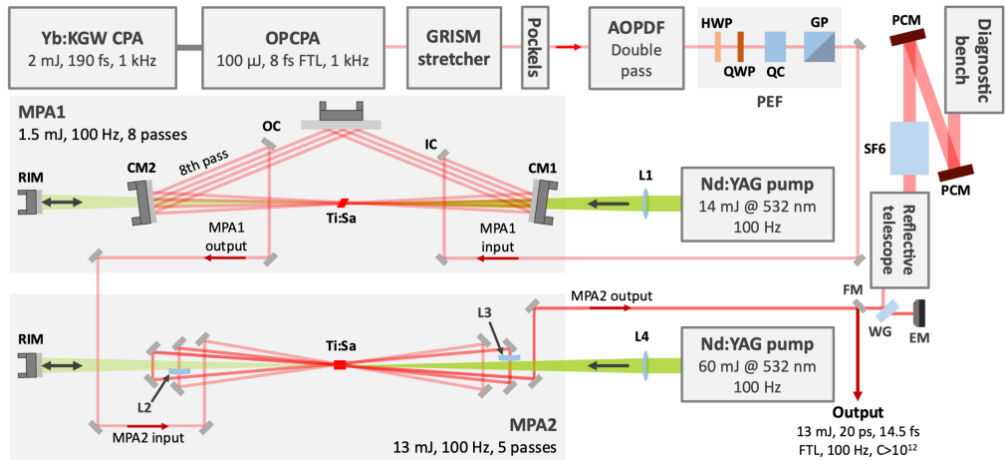

Fig.1. Scheme of the frontend laser system. Ti:Sa MPA1 and MPA2 denotes multipass amplifiers, PEF polarization-encoded filter, HWP and QWP half-wave and quarter-wave plates, QC quartz crystal, GP Glan polarizer, L1-L4 lenses, IC and OC output coupler mirrors, RIM re-imaging mirrors, CM1 and CM2 concave amplifier mirrors, WG wedge, EM energy meter, PCM positively chirped mirrors, FM flip mirror. In MPA1 only four passes are shown for better visibility.

The sampled negatively chirped pulses are recompressed by passing 4 times through a 10 cm thick glass block (Fig. 1, SF6) followed by positively chirped mirrors. Recompression of full energy pulses requires an appropriately sized compressor arrangement installed in vacuum. In this case, based on the measured transmission of the compressor, we calculated an output of 9 mJ at the end of the laser system.

## 2.2 Seeder

Seed pulses for the frontend are obtained via a difference frequency and double-supercontinuum generation scheme, followed by two noncollinear OPA stages amplifying in the spectral region of 680 to 1050 nm [18]. An output energy of 100 µJ is reached with a spectrum corresponding to about 8 fs Fourier transform limited (FTL) duration (Fig. 2). By applying active feedback on the output energy and spectrum, the average power reaches <0.5% RMS stability for 10 hours, while the FTL duration of the spectrum in the same temporal window is kept within 0.23 fs RMS (Fig. 2, a and b). The spatially resolved spectrum of the OPCPA is measured in the near field by an imaging spectrometer calibrated for the wavelength range of 500 to 1500 nm (MISS-SWIR, Femto Easy). The spectrum is optimized so, that most of the pulse energy falls into the gain band of Ti:Sa in the region of 680–920 nm (Fig. 2, c-f). Intensity contrast of the OPCPA seeder is measured after a chirped mirror compressor (PC1607, Ultrafast Innovations, UFI) by using a third-order cross-correlator (Tundra+, UFI). Replica pre-pulses from the SHG process, originating from the post-pulses of the same temporal position, are identified at –12, –30 and –32 ps, (Fig. 6, a). The ragged structure at time before -10 ps is noise typically generated by third order cross-correlators when the energy level of measured signal is substantially lower than that required by the device supplier. Due to the 25 µJ energy available for the measurement and the uncompensated high order dispersion, the dynamic range is limited to about 10 orders of magnitude. As a result of using short pump pulses in the NOPA stages, the pre-pedestal drops to the $10^{-10}$ relative intensity level already before –1 ps.

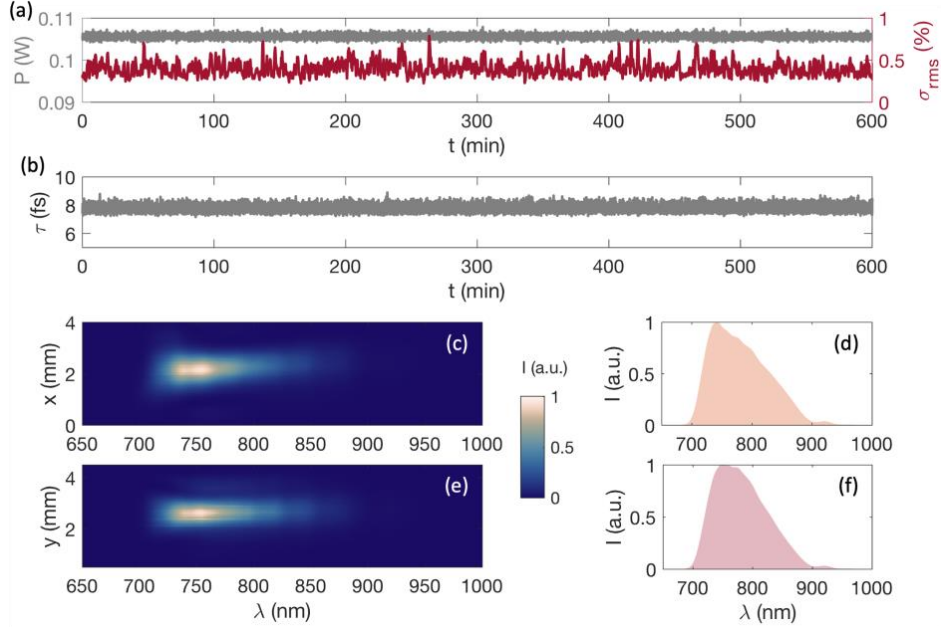

Fig. 2. Performance of the seeder OPCPA: output power and its stability (a), pulse FTL stability (b) over 10 hours, spatially resolved spectrum in the horizontal (c) and vertical (e) cross-sections, and the corresponding spatially integrated spectra (d and f).

Furthermore, the post-pulse pedestal starts at a very low, $10^{-8}$ intensity level, which helps with the attenuation of nonlinear phase accumulation related pre-pedestal growth during the amplification process [17]. Despite the short pump and seed pulses in the OPCPA (<200 fs), at the trailing side of the pulse we observe a long-ragged pedestal, characteristic to CPA systems [19,20]. The underlying reason is subject to future investigation.

### 2.3 Spectral phase and amplitude management

Dispersion introduced by the stretcher must be limited to fall into the dispersion window of the programmable acousto-optic dispersion filter (Dazzler HR650-1050, Fastlite), used in the double-pass mode. Propagation through the Dazzler is set similarly to [13], where back-reflection from a mirror is used after the first pass through the acousto-optic crystal at a small angle, suitable to separate the two passes in the vertical direction. The double-pass operation of the Dazzler offers two advantages: it effectively doubles the dispersion dynamics, while it also compensates most of the spatial chirp introduced by the acousto-optic crystal to the broadband diffracted pulses. In order to utilize the spectral shifting process experienced during chirped pulse amplification in Ti:Sa stages, we use negative stretching in our frontend [20]. As a result, the short wavelength spectral components are more enhanced during amplification. This effect in combination with spectral pre-shaping provides a very efficient way to conserve most of the spectral bandwidth of the seed pulses during amplification.

Due to the designed ultrabroad bandwidth of the amplified pulses, compensation of high order phase distortions becomes even more important. For this reason, we decided to implement a GRISM type stretcher, which gives negative GDD and negative third order dispersion (TOD) and fourth order dispersion (FOD), as well. This allows us to stay in a dispersion window, where material dispersion via propagation through glass can compensate back most of the GDD and TOD at the end of the frontend. The stretched pulse duration is measured by using a cross-correlator arrangement (Light Conversion). The GRISM is set to ensure that the stretched pulse duration is limited to 20 ps at the end of the amplification chain.

Spectral amplitude shaping is obtained by the Dazzler and the PEF stage (Fig. 2, PEF). Here, most of the spectral amplitude manipulation is performed by the PEF, because the spectral phase compensation efficiency with the Dazzler is distorted in case of low spectral amplitudes, which is preferred to be avoided. Nevertheless, narrow spectral amplitude features were tuned with the Dazzler, which helped us to maintain a smooth spectral intensity profile after MPA2.

*2.4 Ti:Sa amplifiers*

Design of our multipass amplifiers is based on numerical simulations, using a home-made spectrally dependent model of amplification in Ti:Sa medium. It also considers the spectral shaping, mirror reflections, polarization-encoded spectral filtering, and the mode matching efficiency between the seed and the pump. By taking parameters of the experimentally available optics, pump lasers, and the seed source, we optimized the amplification in MPA1 and MPA2 for the broadest spectral output. Results of these simulations are presented in Fig. 3 (a-d and j-k). Spectral shaping is more pronounced with the PEF stage, only fine tuning is achieved with the Dazzler. The final spectrum after MPA2 has an FTL of 14.5 fs in combination with an output energy of 13.2 mJ, while according to our modeling, saturation is not reached due to damage threshold limitations imposed by the short stretched pulses in MPA2.

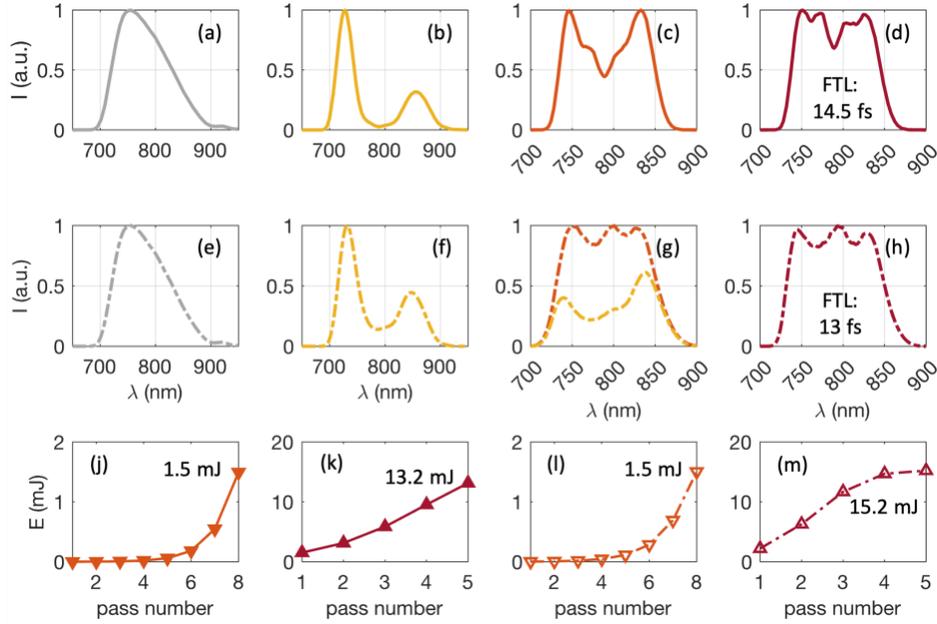

Fig. 3. Simulated spectral evolution in the frontend with the experimentally available optics: seed from OPCPA (a), stretched and spectrally shaped seed (b), MPA1 output (c), and MPA2 output (d). Similar spectral evolution in the frontend with optimized optics: seed from OPCPA (e), stretched and spectrally shaped seed (f), MPA1 output (g, orange) and spectrally shaped MPA2 input (g, yellow), and MPA2 output (h). Energy evolution in MPA1 (j) and MPA2 (k) for the experimental conditions, and MPA1 (l) and MPA2 (m) output for the optimized case.

In comparison, an optimized amplifier system is also simulated with improved coating designs both in terms of spectral reflectance and damage threshold in MPA1 and MPA2. According to our results, an FTL duration of 13 fs in combination with an output energy of >15 mJ can be achieved (Fig. 3, e-h and l-m). This is allowed by using a sequential spectral shaping approach, where we maximize the bandwidth in both MPA1 and MPA2 by using spectral filtering mirrors before and after MPA1. This approach also allows for the mitigation of post-pulse generation in the PEF stage.

In the experimental setup, the mJ-level MPA1 is based on the triangular-ring geometry; thus, it consists of two dielectric coated concave spherical mirrors with a diameter of 50 mm and a focal length close to 500 mm, and a 12 cm long silver coated folding mirror (Fig. 1, MPA1). Pump pulses with 14 mJ energy and 10 ns temporal duration at 100 Hz repetition rate are relay imaged from the pump laser's (Merion-C, Lumibird) output to the central plane of the Ti:Sa crystal. The Brewster-cut Ti:Sa crystal in this amplifier has a path length of 5 mm, which ensures a single pass absorption of 80% for the 532 nm pump pulses. We use a concave dielectric mirror to re-image the transmitted beam to the Ti:Sa crystal to obtain a total absorption level of 96%. In case of optimized spectral shaping with the Dazzler, an output energy of 1.5 mJ is reached with a close to Gaussian spatial profile after the 8th pass. Total gain in MPA1 accounts for about $2.5 \cdot 10^3$ for the seed pulse energy of about 0.6 µJ injected to the amplifier. The Brewster-cut crystal ensures polarization cleanliness and mitigation of post-pulses originating from MPA1. To preserve the spectral bandwidth of the amplified pulses as much as possible, saturation is avoided in this amplifier. At the output of MPA1 a spectrum corresponding to 13 fs FTL duration is achieved, limited by the coatings of the concave mirrors in the amplifier.

Pre-amplified pulses are injected to MPA2 after a mirror telescope, with a small amount of divergence at the entrance of the butterfly arrangement (Fig. 1, MPA2). Pump pulses from a second diode-pumped 100 Hz pump laser (Surelite 4, Continuum) with 60 mJ energy are applied to excite the 6 mm thick, normally cut Ti:Sa crystal with 80% single-pass absorption. The transmitted pump beam is imaged back to the crystal to obtain 96% total absorption. After the second pass, a –4 m focal length lens, and after the 3rd pass a –3 m focal length lens, are incorporated into the beam path to compensate the spatial profile narrowing due to the gain distribution and thermal lensing. Amplification increases the output energy up to 13 mJ with a total gain of 19, while the mode remains close to Gaussian with a size of 1.7 x 1.55 mm at $1/e^2$ intensity level. Due to aiming for a large spectral bandwidth, and the need for spectral phase control, the damage threshold of MPA2 mirrors was found to be a critical parameter. Because of the limited temporal stretching and the close to Gaussian spatial profile of the pump, the required energy density for full saturation of the amplification is not possible to be reached. Without saturation, the stability of MPA2 output energy is below that of the pump laser, i.e. 0.5% RMS shot-to-shot of the pump, and 2.6% RMS on the 1-minute time window of the amplified pulse. In the long term, the amplifier keeps a constant output energy >12.8 mJ after the warm-up time (<5 minutes). By fine tuning the spectral amplitude shaping with the Dazzler, we can obtain a spectrum corresponding to 14.5 fs FTL duration, which matches the value predicted by our numerical simulations.

## 3. Output performance

For testing the pulse recompression, we sample the output pulses of MPA2 by using an uncoated BK7 wedge, and an energy of 1 mJ is sent to the compressor only. Before compression, the output beam of MPA2 is magnified to have a collimated beam size of 11 mm at $1/e^2$ intensity level (Fig. 1, Reflective telescope). The compressor consists of an AR coated 10 cm thick SF6 glass block, through which the pulses propagate four times. A pair of 4" sized positively chirped mirrors (CM1512, Ultrafast Innovations) is used with +100 $fs^2$ dispersion per bounce, and with a total of six reflections for complete dispersion compensation. Transmission of the compressor is measured to be 70%, which would result in 9.1 mJ pulse energy in case of full energy compression. The compressed pulses are resized by a mirror telescope, after which they are temporally characterized by scanning (D-cyce XR, Sphere) and single-shot (D-shot R, Sphere) dispersion-scan devices. Recompressed pulses are found to be as short as 15.5±0.7 fs, depending on the device, while the FTL was 14.5±0.2 fs (Fig. 4, a-d). The measured spectral phase is also shown to be exceptionally stable for the complete spectrum (Fig. 4, e).

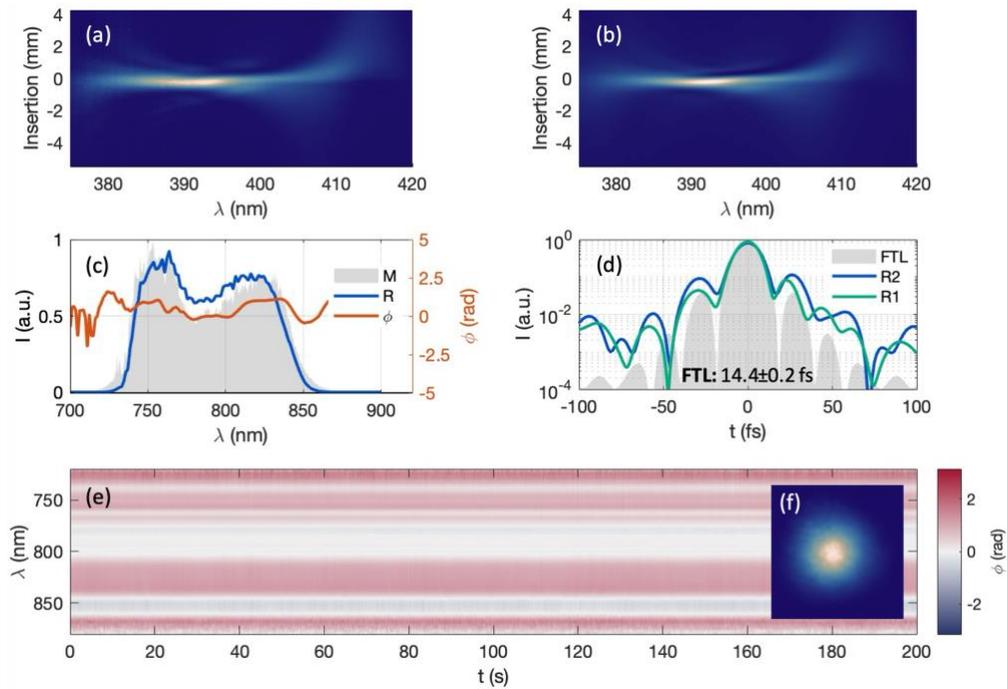

Fig. 4. Measurement results with D-shot R: measured (a) and retrieved traces (b), measured (M) and retrieved (R) spectrum with phase (φ) (c), FTL and retrieved temporal profiles from D-cycle (R1) and D-shot (R2) (d), and spectral phase stability for 1000 shots (e). Spectrally integrated spatial profile after MPA2 is shown in (f).

Spatio-spectral quality of the compressed pulses was investigated by using an imaging spectrometer (MISS-L-B, Femto Easy) and by the ICE (Sphere Ultrafast Photonics, Sphere) [22] technique (Fig. 5). By stitching the temporal characterization results of the D-cycle XR in a small central cross-section of the beam with the spatio-spectral information from ICE, the spatio-temporal profile was reconstructed in both directions (Fig. 5, b and c). The spatio-temporal profile revealed practically zero pulse front tilt and curvature, meaning that the pulse has no residual angular dispersion and chromatic aberrations.

Temporal intensity contrast is measured by using a high dynamic range third order cross-correlator (Sequoia HD, Amplitude), while the contrast of the OPCPA measured with Tundra+ is also presented for comparison (Fig. 6). Our results show that at the laser exit an intensity contrast of $>10^{12}$ is reached already 15 ps before the main pulse (Fig. 6, a and d), which in fact corresponds to the dynamic range of the measurement. The low stretching factor of the seed pulses ensures that the pre-pulses appearing in Fig. 6 at delays before –10 ps are not real. Most of these features are replicas of post-pulses produced by the second harmonic generation process in the third order scanning cross-correlator. This is confirmed by the fact, as visible in Fig. 6, that intensity of pre-pulses marked with B8-B14 equals to the square of that of post-pulses marked with A8-A14. Consequently, the good linearity of the cross-correlator measurement is also shown this way.

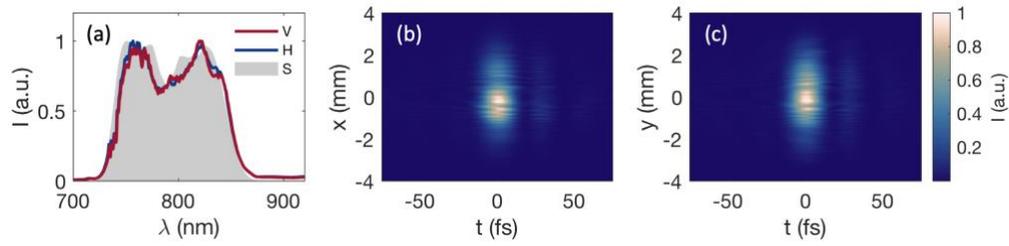

Fig. 5. Spatially integrated spectra in the vertical (V) and horizontal (H) cross-sections, where the simulated MPA2 output spectrum (S) is also shown as a comparison (a). Spatio-temporal profiles are shown in (b) and (c) in the horizontal and vertical cross-sections, respectively.

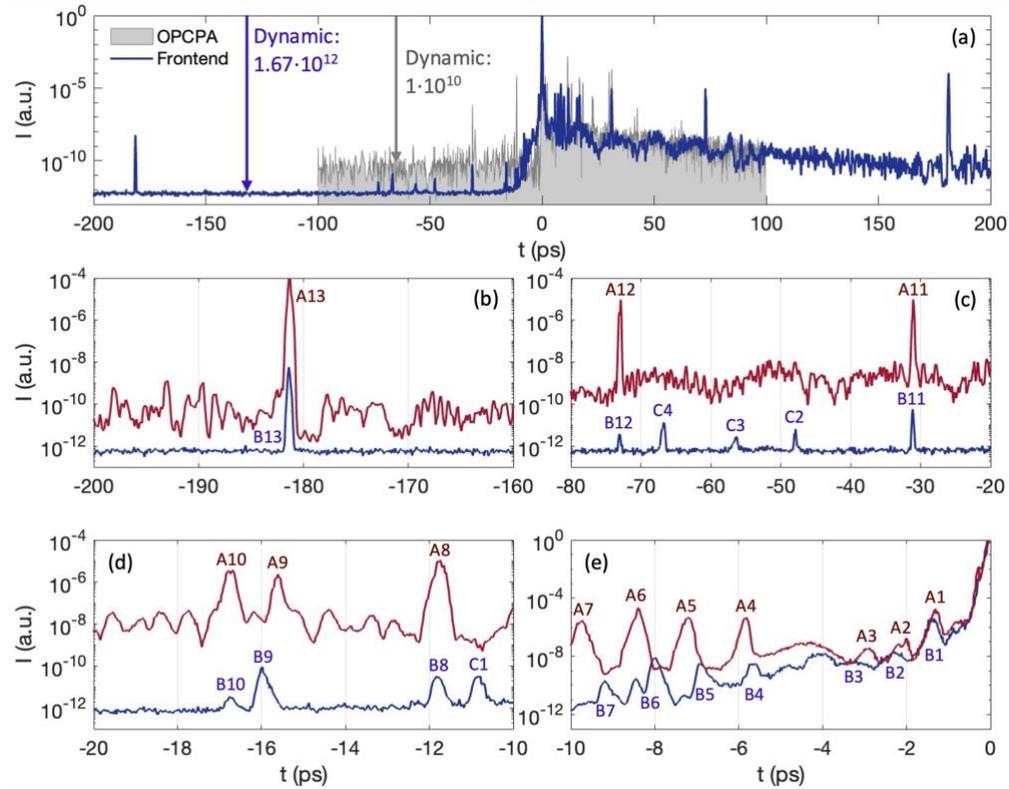

Fig. 6. Temporal intensity contrast on the long temporal range (a), where the contrast of the OPCPA is visualized with grey area, while the output contrast of the complete frontend with blue line. Temporal windows with pre-pulse features are shown in (b) – (e), where the pre-pulse side is drawn with blue lines, while the mirrored post-pulse side is presented with red line for comparison on each plot. Features marked with A are post-, while with B are corresponding pre-pulse pairs. Features marked with C are pre-pulses without post-pulse pairs.

The rising pre-pedestal and pre-pulses visible within the 5 ps vicinity of the main peak, marked with B1-B3, are generated from the post-pedestal and post-pulses originating from optical elements of the OPCPA. In addition, we attribute part of the pre-pedestal to contributions of the GRISM stretcher. Pre-pulses marked with B5-B7 are generated by nonlinear optical coupling [23] from the corresponding post-pulses of A5-A7. Pre-pulses marked with C1-C4 do not have post-pulse pairs in their temporal window, and we identified them as high order ghosts, i.e. THG signals from post-to-post-pulse interactions. Delay mismatch can be also observed in case of several pre-pulses compared to their post-pulse pairs (Fig. 6, d and e), which were attributed to dispersion effects according to a recent study [19,23,24]. Post-pulses marked with A4-A7 are originating from the PEF stage, which could be eliminated by using specially designed, reflection-based polarization manipulation optical elements. In the final state of the development, specifically designed spectrally filtering mirrors could be applied similarly to [13], which can fully eliminate post-pulses. Furthermore, post-pulses marked with A1-A3 can be also eliminated by using wedged optics in the OPCPA. It is remarkable that at the system exit the ragged post-pedestal starts at the same level as it appears after the OPCPA seeder (Fig. 6, a). It starts at the $10^{-8}$ relative intensity level, which is lower than what is typically observed in Ti:Sa lasers. This is very advantageous in case of further amplification in a consecutive CPA chain for keeping the pre-pedestal generation under control via nonlinear phase accumulation.

## 4. Discussion

Scaling of the output amplified spontaneous emission (ASE) contrast of DCPA-based laser systems shows a linear dependence on the stretched pulse energy available at the input of the first amplifier, assuming, that the seed has negligible ASE. This effect was thoroughly investigated in [15], where the linear dependence on the input energy was identified. Together with the original results from [15], we plot results reported in high contrast laser systems in Fig. 7. The dramatic increase in achievable temporal intensity contrast before the main pulse for elevated input pulse energies makes a clear indication on the requirement of high energy seed for future frontend architectures. The inherently high ASE contrast of a frontend is evidently a major aspect, which makes a solid argument for OPCPA seeders (Fig. 7, ELI ALPS HFPW frontend [11]).

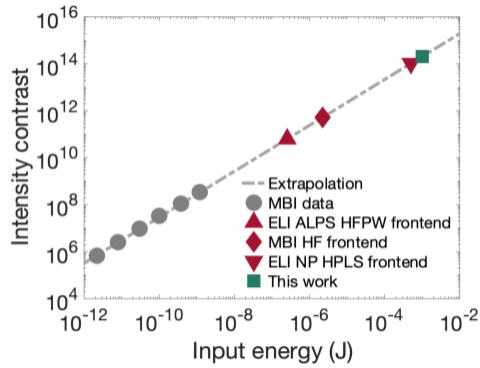

Fig. 7. ASE contrast limitation for different injected pulse energies into the first Ti:Sa amplifier after the main stretcher. Grey dashed line is extrapolation based on results visualized by grey dots [15].

The energy of the light seeded into the amplification chain is limited currently to a few millijoules. It is given by the damage threshold of diffraction gratings of the stretcher and Dazzler, which is typically used for the fine tuning of the spectral phase. In addition, most of currently used stretcher types possess a spatially dependent group delay that does not allow to use beams with an aperture of more than 2 mm. With seed energies in the multi-µJ range the achievable ASE contrast is close to $10^{12}$ (Fig. 7, MBI HF frontend [16]). If the seed pulse has both the $>10^{14}$ ASE contrast ratio and input energy of a mJ level, the ASE contrast at the output of the amplifier chain can be as high as $10^{14}$, which is extremely important for experiments at the 10 PW level (Fig. 7, ELI NP HPLS frontend [10]). The frontend architecture described in this work supports the above-mentioned requirements based on our extrapolation (Fig. 7, This work), where a seed energy of 10 mJ at the input of the stretcher is taken in combination with negative stretching.

Compared to XPW-seeded picosecond OPCPA [10], or fully OPCPA-based architectures providing multi-mJ energy before the main stretcher, our design combines the advantages of both OPCPA and Ti:Sa-based amplification. The seeder OPCPA provides the bandwidth, high gain amplification from nJ to 100 µJ energy, ultrahigh ASE contrast and day-to-day reproducibility. In the meantime, the spectrally shaped and negatively stretched Ti:Sa-based amplification can be driven by nanosecond, highly stable Nd:YAG DPSSL pump lasers, providing simplicity and robustness against picosecond jitter. Since our frontend does not require a broadband mode-locked oscillator, nonlinear spatio-temporal cleaning, or picosecond pump lasers, it also presents a cost-effective solution.

## 5. Conclusions and outlook

In conclusion, here we demonstrate a hybrid frontend scheme, which is able to deliver 13 mJ pulses at 100 Hz with sub-15 fs FTL duration, excellent spatio-spectral quality and ultrahigh temporal contrast. The architecture has the potential for increasing the pulse energy above 15 mJ in combination with a spectrum corresponding to 13 fs FTL, and further optimized temporal contrast without pre-pulses in the 10 ps vicinity of the compressed pulse. Advanced performance of our frontend is obtained thanks to the following key design points.

Industrial grade OPCPA seeder with 100 μJ energy, which provides high stability, ultrabroadband spectrum, and ultrahigh temporal contrast due to the double WLG scheme and the femtosecond NOPA stages.

Applying a scheme with negative stretching and positive compression supports blue spectral shifting of the amplified spectrum in the CPA method, which is preferable for further amplification in power amplifiers with positive pulse chirping [15]. Recompression with bulk allows for low energy losses.

We use an advanced spectral shaping arrangement consisting of a combination of AOPDF and polarization-encoded spectral filter, which allows for a fast and flexible tuning of the amplified spectrum, while keeping associated energy losses at a moderate level.

By limiting the stretched pulse duration to 20 ps, we minimize the temporal window where pre-pulses and pre-pedestal can grow via accumulation of nonlinear phase during amplification of chirped pulses [29].

As a result of the above-mentioned design features, an ultrahigh temporal contrast of $>10^{12}$ is achieved without using any nonlinear spatio-temporal cleaning and picosecond OPCPA stages.

Bandwidth of the amplified pulses in MPA1 is still limited by the amplifier mirror coatings, which are to be further improved, reaching sub-12 fs FTL pulses. Implementation of sequential spectral shaping by using filtering mirrors can further improve the spectral evolution inside MPA1 and MPA2, allowing for lower intensity of the stretched pulses during amplification. This would in turn enhance the temporal contrast even further with less nonlinear phase accumulation. Improvement of the energy stability is planned by using mirrors with a higher damage threshold and improved pump mode quality in the Ti:Sa crystal in MPA2, pushing this stage into gain saturation.

Compared to other currently available frontend architectures, our system can provide a significantly enhanced reliability due to the industrial grade OPCPA seeder, which produces seed pulses with a temporal contrast of $>10^{13}$ at time preceding few picoseconds. The amplification stages are driven by industry standard nanosecond Nd:YAG DPSSLs, which do not require optical seeding from a master oscillator. This makes the amplification section of the system resilient to temporal jitter in the picosecond temporal domain, compared to ps-OPCPA stages, which are implemented in [1] and [10]. By using the combination of PEF and the AOPDF, the output spectrum can be pre-shaped to the requirements of further amplification. With the ultrabroad bandwidth demonstrated in this work, the pre-shaping capability gives the possibility to obtain sub-17 fs pulses at the end of 100 TW-class amplifier chains. In this case, the 20 ps stretched output pulses of the frontend could be directly seeded into a stretcher, which allows to reach a current limit on the final ASE contrast ratio of a PW-class amplifier chain above $10^{14}$ (Fig. 1). Furthermore, the energy at the output of our frontend is sufficient to seed two PW-class amplifier chains simultaneously.

**Funding.** European Regional Development Fund (GINOP-2.3.6-15-2015-00001).

**Disclosures.** The authors declare no conflicts of interest.

**Data availability.** Data underlying the results presented in this paper are not publicly available at this time but may be obtained from the authors upon reasonable request.